# Excitation of surface plasmon polariton modes with multiple nitrogen vacancy centers in single nanodiamonds

Shailesh Kumar[1,*,†], Jens L. Lausen[1,†], Cesar E. Garcia-Ortiz[2], Sebastian K. H. Andersen[1], Alexander S. Roberts[1], Ilya P. Radko[1], Cameron L. C. Smith[3], Anders Kristensen[3] and Sergey I. Bozhevolnyi[1]

[1] Centre for Nano Optics, University of Southern Denmark, DK- 5230 Odense M, Denmark
[2] CICESE, Unidad Monterrey, Alianza Centro 504, PIIT Apodaca, Nuevo Leon, 66629, Mexico
[3] Department of Micro- and Nanotechnology, Technical University of Denmark, DK-2800 Kgs. Lyngby, Denmark

[*] *Corresponding author: shku@iti.sdu.dk*
[†] *These authors contributed equally to this work.*

**Abstract:**

Nitrogen-vacancy (NV) centers in diamonds are interesting due to their remarkable characteristics that are well suited to applications in quantum-information processing and magnetic field sensing, as well as representing stable fluorescent sources. Multiple NV centers in nanodiamonds (NDs) are especially useful as biological fluorophores due to their chemical neutrality, brightness and room-temperature photostability. Furthermore, NDs containing multiple NV centers also have potential in high-precision magnetic field and temperature sensing. Coupling NV centers to propagating surface plasmon polariton (SPP) modes gives a base for lab-on-a-chip sensing devices, allows enhanced fluorescence emission and collection which can further enhance the precision of NV-based sensors. Here, we investigate coupling of multiple NV centers in individual NDs to the SPP modes supported by silver surfaces protected by thin dielectric layers and by gold V-grooves (VGs) produced via the self-terminated silicon etching. In the first case, we concentrate on monitoring differences in fluorescence spectra obtained from a source ND, which is illuminated by a pump laser, and from a scattering ND illuminated only by the fluorescence-excited SPP radiation. In the second case, we observe changes in the average NV lifetime when the same ND is characterized outside and inside a VG. Fluorescence emission from the VG terminations is also observed, which confirms the NV coupling to the VG-supported SPP modes.

## 1. Introduction

Nitrogen-vacancy (NV) centers in diamonds have attracted much attention as a possible candidate for solid-state quantum bits (qubits) [1, 2]. Their ground-state electron-spin coherence time is sufficiently longer than the time it takes to perform a qubit operation in such qubits. The electronic qubit state can be initialized and read out optically and, importantly, even at room temperatures [3-6]. The fluorescence from NV centers is stable, and a single NV center is a stable source of single photons [7]. NV centers are also a promising candidate for a high-precision magnetic field, electric field and temperature sensors [8-14]. It is argued that high densities of NV centers are needed for achieving the best performance of NV centers in magnetic field sensors. Multiple NV centers in nanodiamonds (NDs) can provide high sensitivity (depending on number of NV centers in the ND) as well as high resolution (determined by the size of the ND) for the magnetic field sensors [11, 12]. In addition, NDs with multiple NV centers are also useful as a stable fluorophore in biological applications due to their chemical neutrality and brightness [15]. In general, the applications potentially enabled by NV centers stand to benefit from the enhancement of the fluorescence rate as well as the efficient collection of the fluorescence from the NV centers. There are several methods that have been proposed and investigated in this direction, such as coupling NV centers to diamond pillars [16], dielectric waveguides [17], dielectric cavities [18-20], plasmonic nanostructures [21, 22] and plasmonic waveguides [23-28].

Waveguide configurations supporting the SPP propagation, i.e., plasmonic waveguides, exhibit (at least, in some cases) a unique feature of guiding SPP modes with extreme confinement, far beyond the diffraction limit [29]. This opens a way to realize a very efficient coupling of emitters to these SPP guided modes [30-32]. Channel plasmon polariton (CPP) modes supported by VGs cut into metal are especially suited for efficient coupling to emitters due to their strong confinement and relatively long propagation lengths [28, 33, 34]. The so-called $\beta$-factor, defined as the probability of a spontaneous decay of an emitter leading to excitation of the plasmonic waveguide mode, is not only high for CPP modes but also distributed more uniformly in space (within a VG), compared to other plasmonic waveguides such as cylindrical and wedge waveguides [33]. VGs have been proposed for realization of long-distance resonant energy transfer and super-radiance effects [33] as well as long-distance entanglement of two quantum emitters [34]. In addition, a wide range of devices, including plasmonic circuit components and nanofocussing elements formed by VGs, have also been demonstrated [35-37]. With nano-mirrors fabricated at the VG terminations, the efficient in- and out-coupling of far-field propagating radiation and CPP modes is demonstrated to be feasible [38, 39].

In this work, we investigate the possibility of exploiting multiple NV centers in individual NDs for coupling to the surface plasmon polariton (SPP) modes supported by silver surfaces (protected by thin dielectric layers) and by gold V-grooves (VGs) produced via the self-terminated silicon etching. This article is organized as follows: We first consider, in section 2.1, the coupling of multiple NV centers in NDs to the SPP modes supported by a silver-dielectric interface. The SPPs propagating along the interface are scattered by another ND. We conduct measurements of the fluorescence spectra obtained from a source ND, which is illuminated by a pump laser, and from a scattering ND illuminated only by the fluorescence-excited SPP radiation. The normalized fluorescence spectrum (by the source spectrum) measured from the scattering ND shows short-wavelength attenuation, which we explain by using an analytic expression based on the point-dipole approximation. Next, in Section 2.2, we consider the coupling of multiple NV centers in an ND to the CPP mode supported by a gold VG, which is fabricated using UV lithography and self-terminated silicon etching [39]. The investigated ND is pushed from the gold-air interface outside the VG to inside the VG using an atomic force microscope (AFM) [22, 24-28]. Fluorescence lifetimes are measured before and after movement of the ND, revealing a noticeable change in the average lifetime that we associate with the NV-CPP coupling. Emission from the VG termination mirrors, when the ND is excited, further confirms this coupling. We terminate our paper with Section 3, in which we summarize the results obtained and offer our conclusions.

## 2. Experiments and Results

The experiments are performed using NDs with an average diameter of 100 nm and containing ~400 NV centers each (Adámas Nanotechnologies). To characterize our systems, we use a scanning confocal microscope. A 532-nm linearly polarized pulsed laser with pulsewidth of ~50 ps and a repetition rate varying between 2.5MHz and 80 MHz is used as the excitation source for NV centers in NDs. The laser beam is focused with a 100x objective (NA 0.9) onto the sample to a spot size of ~500 nm. The same objective is used for collection of the fluorescence signal. A long pass filter at 550 nm filters out the signal from the excitation source. The fluorescence is either temporally resolved with an avalanche photo diode (APD) and counting electronics, or spectrally resolved with an electron-multiplying charge-coupled device (EMCCD) camera mounted on a spectrometer. A schematic of the set-up is presented in Figure 1(a).

## 2.1 Coupling of NV centers in an ND to SPPs at a silver-dielectric interface

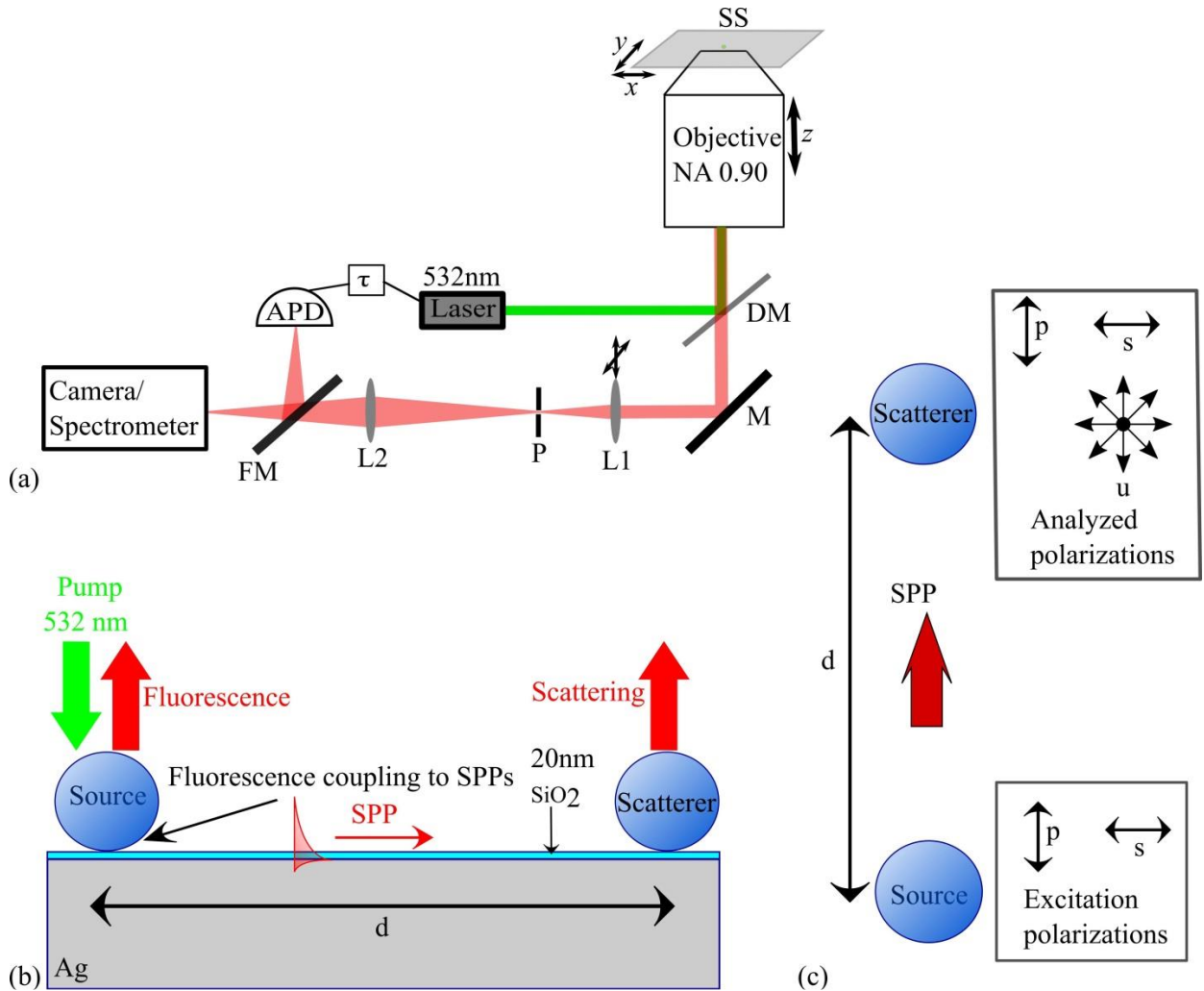

*Figure 1: (a) Schematic of our experimental set-up. SS: sample stage, DM: dichroic mirror, M: mirror, L1 and L2: lenses, P:pinhole, FM: flip mirror, APD: avalanche photodiode, and τ denotes the timing electronics. The arrows indicate the possibility of moving the corresponding components. (b) side view and (c) top view of the experimental set-up for the NDs on a silver/glass surface. A pump laser excites NV centers in the source ND, where some of the fluorescence couples to the SPP mode and the scatterer ND scatters the SPPs to the far field. The source ND is excited by polarizations parallel to the line joining the source and the scatterer NDs (p) and perpendicular to the line (s). The signal is analyzed in these two polarization directions as well as unpolarized (u).*

The sample for studying the excitation of SPPs propagating along a silver-dielectric interface is prepared by thermal evaporation of an optically thick 150 nm silver film on a silicon wafer. Subsequently, a 20 nm amorphous $SiO_2$ film is sputtered on top of the silver film while it remains in vacuum in order to minimize its reaction with atmospheric sulphur and oxygen. NDs are deposited on the sample using a technique that has been described in [40]. Briefly, spin coating of PMMA [Poly(methyl methacrylate)], electron-beam lithography and development is used to fabricate an array of holes in PMMA. A suspension of NDs in water is then applied, is left to dry out under ambient conditions, thereafter a lift-off is used to remove the PMMA and obtain an array of NDs on the sample surface. The process ensures that the NDs are placed periodically at a controllable distance. The experiments are performed for two different distances between NDs: 7 and 9 µm.

The excitation polarization is changed between polarization parallel to the straight line connecting the source and the scattering ND (referred to as *p* polarization) and that perpendicular to the line (referred to as *s* polarization), as indicated in Figure 1(c). Likewise, the signal from the NDs is analyzed in these two different orientations (*p* and *s*) and also in the absence of analyzer (*u*), which is also indicated in Figure 1(c).

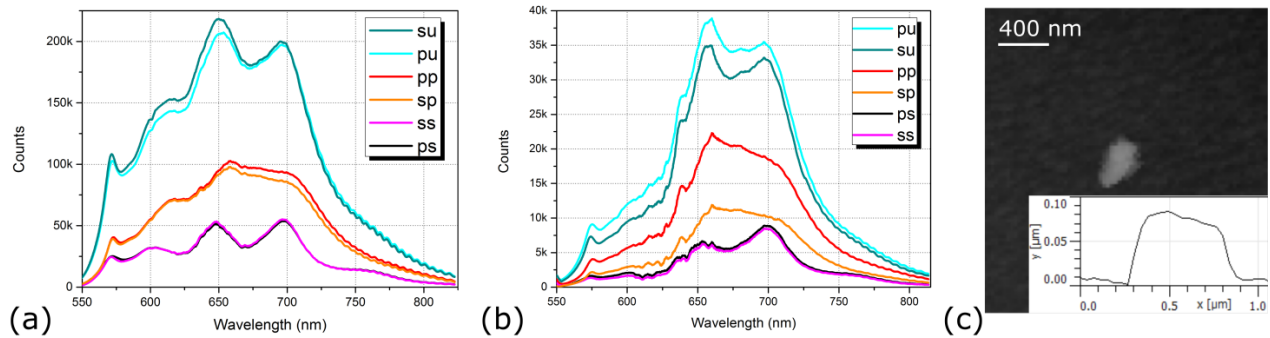

*Figure 2: Typical spectra measured at the source (a) and scatterer (b) ND with different combinations of excitation and analyzed polarizations. In the graphs, the first letter denotes excitation polarization, whereas the second letter denotes analyzed polarization. (c) shape of a single ND imaged with an AFM. The inset shows a cross-section of the ND.*

When an ND is excited, its fluorescence couples to SPPs at the silver-dielectric interface, propagates along the interface and is scattered by a separate ND. In Figure 2(a) and 2(b), we present the source and the scattering spectra collected from the NDs, which are separated by 7 µm. Different configurations of polarization orientations for excitation radiation and analyzed signal are indicated in the insets in the respective figures. No significant difference in the strengths of signals collected from the source ND for the two orthogonal excitation polarizations is observed, a feature that can be explained by a high number and arbitrary orientation of NV centers inside the ND. The spectrum is found to depend only on the analyzed polarization, which we attribute to an irregular shape and relatively large size of the source ND [Figure 2(c)] resulting in complicated Mie scattering spectra. For other source scatterer ND combinations, we have observed fluorescence dependence on excitation polarization (data not shown).

The signal from the scatterer ND is found dependent on the excitation polarization [Figure 2(b)]. The unpolarized collected signal for the *s*-polarized excitation (su) is lower compared to the unpolarized collected signal for the *p*-polarized excitation (pu). The same is observed when an analyzer is present: excitation by the *p*-polarized light consistently gives higher signals from the scatterer ND (the difference is small for the *s*-analyzed orientation). The dipoles excited by *p* polarization couple preferentially to the SPP propagating in the direction along the electric field of excitation [40]. A signal is still observed for the *s* polarization excitation, because many of the excited NV dipoles have a nonzero projection on the line connecting the source and scatterer NDs and can therefore couple to the SPPs propagating in that direction. In general, the relative strength of ND-excited SPP radiation depends on the distribution of dipoles inside the source ND [41]. Note that, similarly to the source ND spectra, the shape of the scatterer ND spectra is determined by the analyzed polarization [Figure 2(b)], a feature that can again be related to the ND shape and size influence on the scattering spectra.

The signal strength collected from the scatterer ND is, in general, significantly lower compared to that collected from the source ND. Comparing the overall (over the whole spectrum) photon counts from the scatterer ND with that from the source ND, we found that their ratio varies substantially from one pair of NDs to another. It cannot be explained merely by variations in ND separation due to imperfect fabrication. The largest ratio observed is 1:2850 and the lowest is 1:4 (data not shown), where these two different pairs of NDs have the same separation of 7 μm. The decrease in signal strength when changing the detection from the source to scatterer ND is therefore not only related to the SPP propagation loss (as conjectured previously [40]) but also to the excitation polarization state, SPP coupling and scattering efficiencies, ND shape, and NV dipole-to-interface distance. We note that the ratio of scattered intensities pp:ps and sp:ss should, in principle, be the same (as observed for a distance 24 µm between source and scatterer [40]) . When the distance is smaller (7 and 9 µm), a small background due to the direct illumination of scatterer (by pump or fluorescence, which affect the ps and ss spectra the most) changes these ratios.

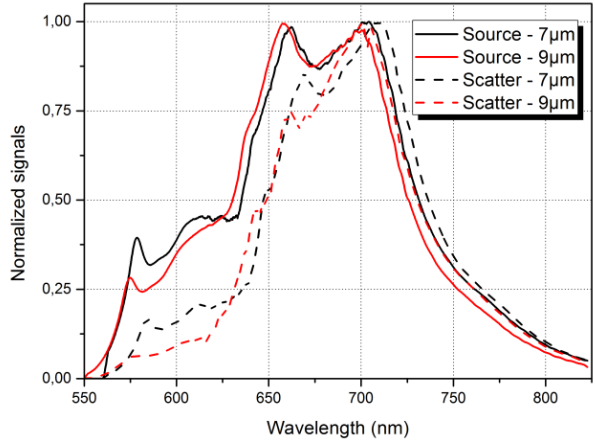

*Figure 3: Source and scatterer spectra, shown for two different distances (7 and 9 μm) between the NDs. All signals are normalized to their own maximum value.*

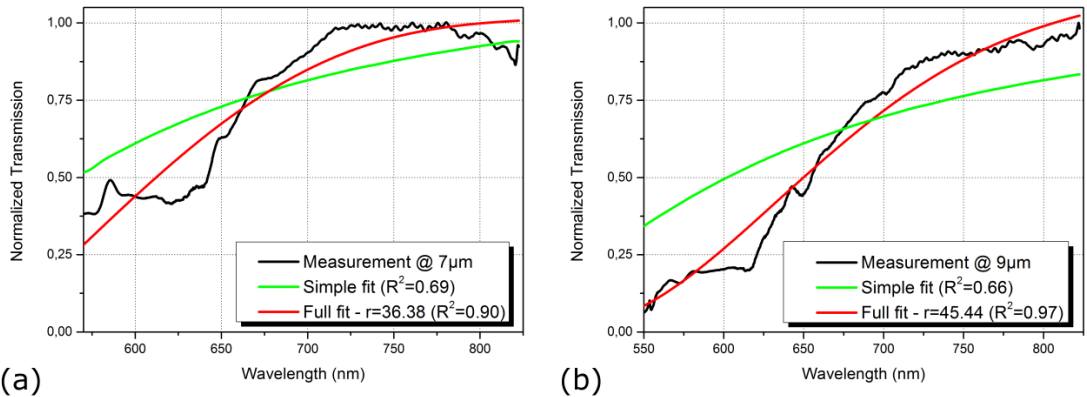

*Figure 4: (a) and (b) shows model fits of the normalized source and scatterer spectra ratios obtained for distances between source and scatterer of 7 μm and 9 μm, respectively. The fits with the simple model, as well as the full model containing terms for divergence and coupling of fluorescence to SPP are shown. The fits for the full model are shown in the legends of the graphs with a fit $R^2$ coefficient.*

We use two different analytical models to fit the ratio of spectra measured from the source and the scatterer NDs. In the first model suggested previously [40], it is assumed that the ratio of spectra can be expressed as follows:

$$\frac{P_{scat}(\lambda)}{P_{source}(\lambda)} = f(\lambda)\, exp[-d/L_{spp}(\lambda)] \ \text{with}\ f(\lambda) = \frac{\eta_{spp}(\lambda)\sigma_{sc}}{2\pi\eta_{rad}(\lambda)da_s}, \qquad (1)$$

where $d$ is the distance between the NDs, $L_{spp}(\lambda)$ is the SPP propagation length for a given wavelength $\lambda$, $\eta_{spp}(\lambda)$ is the photon to SPP coupling efficiency for the source ND, $\sigma_{sc}$ is the scattering cross-section of the scatterer ND, $\eta_{rad}(\lambda)$ is the photon emission efficiency of the source ND, and $a_s$ is the in-plane angular dependence of the SPP intensity. In this rather simple model, the term $f(\lambda)$ is assumed to be wavelength-independent, i.e., taken as a constant, which is the fitting parameter for the model. Since $f(\lambda)$ is used as a multiplication factor, all spectra are normalized to their own maximum. In the second model, we have attempted to develop a more accurate and yet analytic description by using the expression for the SPP contribution to Green's dyadic constructed for evaluating SPP-mediated interaction between small particles located close to a metal surface [42]. When applied to our configuration, one arrives at the following expression:

$$G_{spp}^{zz}(d,\lambda) = \frac{k_{spp}}{2\sqrt{\varepsilon}}\sqrt{\frac{2}{\pi k_{spp}d}}\, e^{-2r\sqrt{k_{ND}^2-k_0^2}}\, e^{-(d/L_{spp}(\lambda))}, \qquad (2)$$

where $k_{spp}$ is the magnitude of the wave vector for the SPP at the silver-dielectric interface, $k_{ND}$ is the magnitude of the wave vector at the silver-diamond interface, $r$ is an average radius of NDs, $\varepsilon$ is the dielectric constant of silver, $L_{spp}(\lambda)$ is the SPP propagation length at the silver-dielectric interface, and $d$ is the distance between the two NDs. The first exponential term accounts for the SPP excitation (by a dipole source, such as an NV center) and out-coupling (by a dipole scatterer) to free propagating radiation, and the second exponential term describes the SPP propagation loss. The square root term describes the angular divergence of the propagating SPP. Using the expression for Green dyadic above (source and scatterer dipole are assumed to have z-orientation), we write the ratio of detected intensities as $P_{scat}(\lambda)/P_{source}(\lambda)=\alpha|G^{zz}_{spp}(\lambda,d)|^2$, where $\alpha$ is a fitting parameter constant for the second model (together with the ND radius $r$). The value of $k_{spp}$ in Equation (2) for a three-layer structure (air/glass/silver), and $k_{ND}$ for a diamond-silver interface (the air above and the glass beneath the ND are neglected) are obtained analytically. The normalized scattering spectra for 7 and 9 μm of separation between NDs (Figure 3) appear shifted to longer wavelengths relative to the fluorescence spectra measured from the source NDs. The SPP attenuation is larger for shorter wavelengths, which contributes to the spectral shape of the scattered spectrum. The coupling efficiency of fluorescence to SPPs and angular divergence of the SPP beam depend on the wavelength, both of which also influence the shape of the scattered spectrum. This is readily seen in Figure 4, where the extended model provides better fitting. The best fit for the extended model has the coefficient of determination ($R^2$) of 0.90 and 0.97 for the distances of 7 and 9 μm, respectively. The ND radii obtained from fits, $r = 36.38$ nm for the distance of 7 μm and $r = 45.44$ nm for the distance of 9 μm, correspond well with the NDs that are actually deposited (an average diameter of 100 nm).

## 2.2 Coupling of NV centers in a ND to CPP modes in a VG

The VGs used in these experiments are fabricated with the same procedure as described in reference [39] which consists of UV lithography, crystallographic etching of silicon and a thermally grown silicon dioxide ($SiO_2$) layer on the silicon to modify the V-shape geometry. Initially, a 200-nm $SiO_2$ layer on the silicon substrate is patterned by both UV lithography and reactive-ion etching to define the perimeter of the VG devices. It is necessary that the patterning of the $SiO_2$ layer is well-aligned with the crystal ⟨100⟩ planes of the silicon substrate. The VGs and termination mirrors are formed by anisotropic wet etching of the exposed silicon in a potassium hydroxide (KOH) bath at 80 °C. The KOH etch yields smooth ⟨111⟩ VG sidewalls and termination mirrors with a fixed inclination of 55° from the surface plane. The nature of the crystallographic etching yields excellent mirror formation at the ends of the VG perimeter. Tailoring of the V-shape geometry is performed by thermal wet oxidation of the silicon VGs (1150 °C for 9 h, resulting in a 2320 nm $SiO_2$ layer at flat sections of the substrate) which sharpens the interior angle of the groove in order to support a CPP mode. The metal is deposited by electron-beam evaporation: first a 5 nm layer of chromium to promote adhesion before a 70 nm layer of gold. The gold layer is chosen to be sufficiently thick to eliminate interaction of air−interface plasmons with the underlying $SiO_2$ layer and also to minimize self-aggregation.

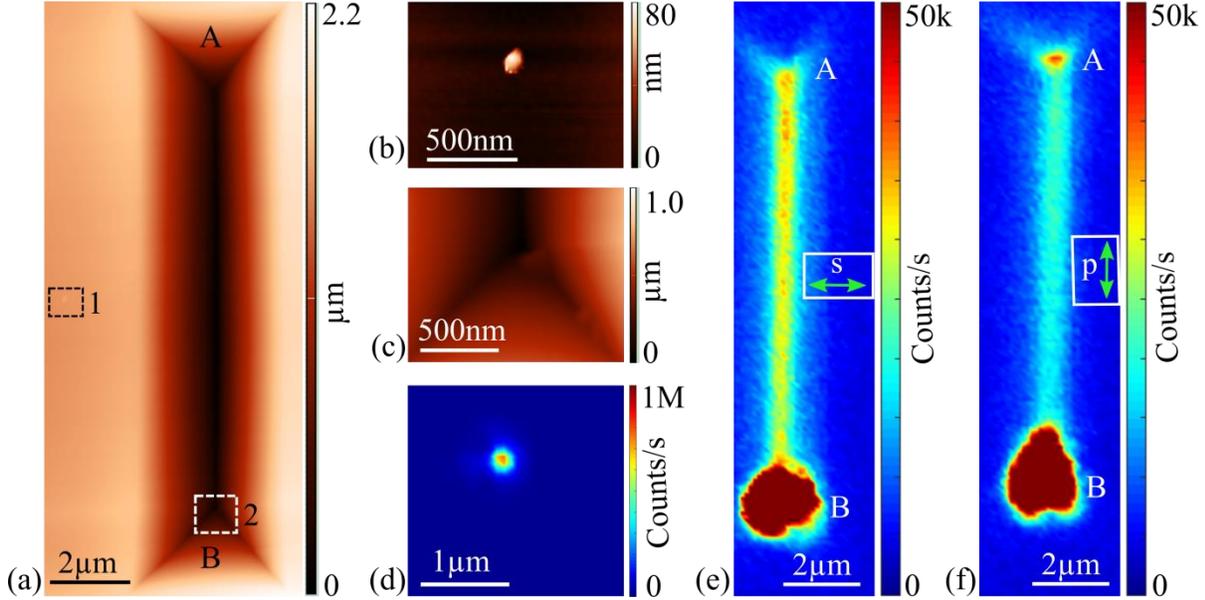

*Figure 5: (a) Atomic-force microscope image of a VG structure. The rectangles with dotted lines indicates the areas containing NDs. (b) and (c) Zoomed-in images of the areas 1 and 2, respectively, indicated in (a). (d) Confocal-microscope fluorescence scan of the area containing the ND presented in (b). (e) and (f) Confocal-microscope fluorescence scans of the VG presented in (a) with the excitation polarization indicated in the inset of respective figures. In (a), (e) and (f), A and B indicate the two VG termination mirrors.*

To couple NDs to the VGs, a water suspension containing NDs is spin coated on the sample with VGs. For this experiment, the concentration of NDs in water is adjusted so that the density of NDs on the surface after spin-coating is less than 1 per 10x10 µm$^2$. An atomic force microscope (AFM) image of a VG together with an NDs is presented in Figure 5(a). As can be observed from the figure, the depth of the VG structure is more than 2 µm. Figures 5(b) and 5(c) show images of the areas containing the NDs, which are also indicated in 5(a). A confocal raster scan image of the area with the ND presented in Figure 5(d), shows that the fluorescence from the ND on the flat gold-air interface at an excitation of 100 µW is ~700 kcounts/s. Confocal scan images taken at an excitation power of 100 µW and two orthogonal polarizations show [Figure 5(e-f)] that fluorescence from the VG is higher in case of excitation polarization across the VG [*s* polarization in Figure 5(e)] due to the contribution from gap plasmons [43]. In Figure 5(f), termination mirror A can be seen to fluoresce more than it does in Fig. 5(e), this is due to the structure of the termination mirror which allows the penetration of p-polarized excitation light and the fluorescence is enhanced due to the presence of three corners of the V-groove termination. The termination mirror B of the VG fluoresces more due to the presence of NDs in the region, which can be seen in Figure 5(c).

The ND shown in Figure 5(b) is further characterized by measuring its lifetime and fluorescence spectrum. The measured lifetime curve as well as a two-exponential fit (Aexp(-t/$\tau_A$) + Bexp(-t/$\tau_B$)+constant, where A and B are amplitudes, t is time, $\tau_A$ and $\tau_B$ are decay lifetimes) is presented in Figure 6(a). We note that NV-centers in nanodiamonds are known to have a distribution in decay lifetimes [23]. Our NDs have ~400 NV centers and each of the NV centers may have different lifetimes. Here, we have fitted the experimental data with minimum number of exponential decay rates possible. A single decay rate does not fit the data well, and a two exponential decay curve fits the data quite well (0.9<$R^2$<1.1). The first 2 ns of the decay curve are not included in the fit to exclude gold fluorescence [44]. The two lifetimes obtained from the fit are $\tau_A$ =13.86 ns and $\tau_B$=5.18 ns. These lifetimes are shortened when compared to those obtained when the NDs from the same suspension are spin-coated on a fused silica substrate. In the latter case, we obtained $\tau_A$ =34.51 ns and $\tau_B$=7.77 ns (averaged over 10 NDs). The difference in lifetimes can be attributed to the excitation of SPPs on the gold-air interface by the NV centers. The fluorescence spectrum of the ND clearly shows [Figure 6(b)] the characteristic zero phonon lines of $NV^0$ (575 nm) as well as $NV^-$ (637 nm) centers.

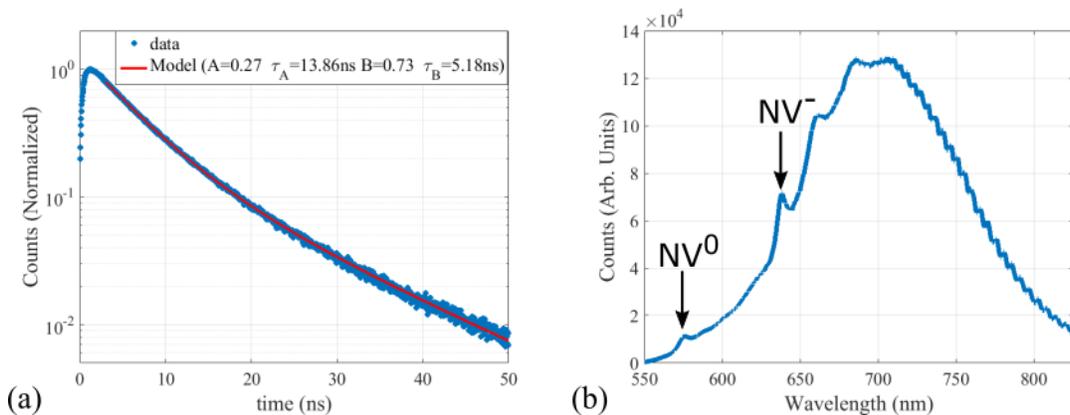

*Figure 6: (a) Lifetime measurement data and a two-exponential fit for the fluorescence from the ND presented in figure 5(d). (b) Emission spectrum of the ND shown in figure 5(d).*

After characterization of the VG as well as the ND, the ND is moved into the VG using a procedure that has been used previously [22, 24-28]. An AFM image of the VG containing the ND can be seen in Figure 7(a). Figure 7(b) shows a zoomed-in image of the ND within the VG, and it can be inferred that the ND is close to the bottom of the VG. The cross-section along the dotted line in Figure 7(b) is presented in Figure 7(c). The height of the ND inside the VG is around 90 nm, which is similar to the height observed outside of the VG for the ND.

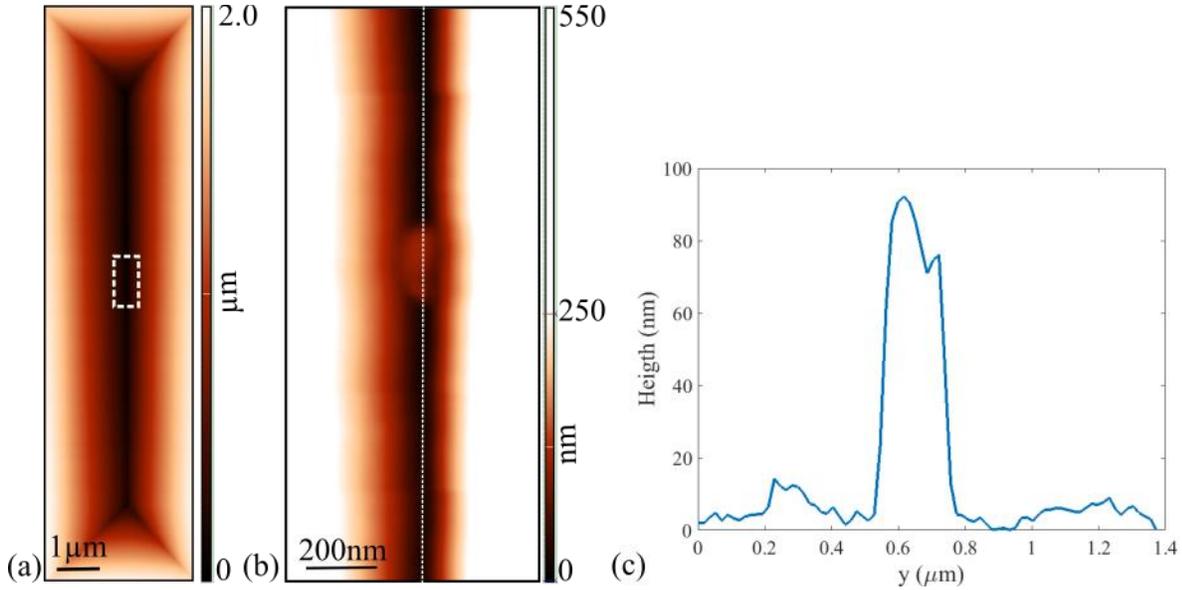

*Figure 7: (a) AFM image of the VG after the ND is moved inside. The white rectangle indicates the area containing the ND. (b) Zoomed-in image of the ND inside the VG. (c) The cross-section along the dotted white line in figure (b).*

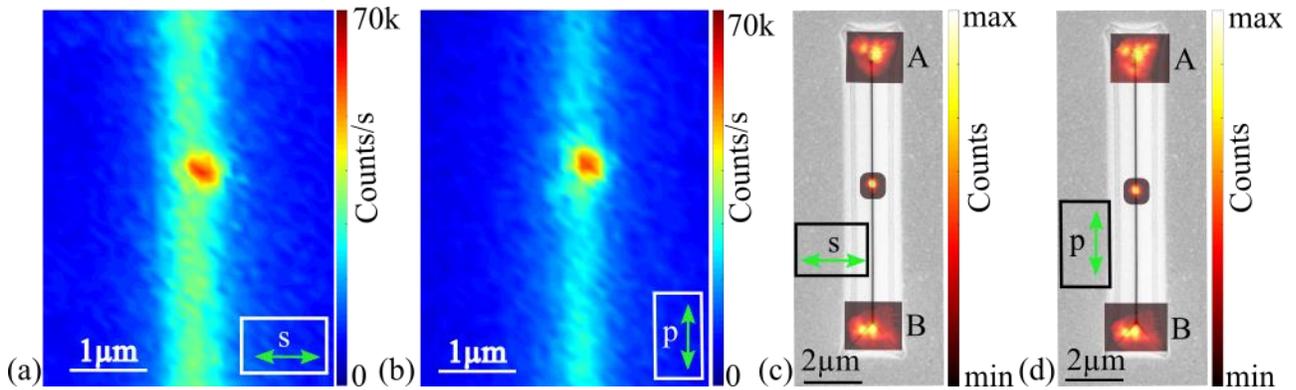

*Figure 8: (a), (b) Confocal microscopy scan images of the ND located inside the V-groove with the excitation polarization indicated in the insets. (c) and (d) Fluorescence images combined with SEM images indicating the emission from the far ends of the VG. The excitation spot is positioned at the ND and the excitation polarization is indicated in the insets. The three spots in both images were recorded separately. A and B indicate the two VG termination mirrors.*

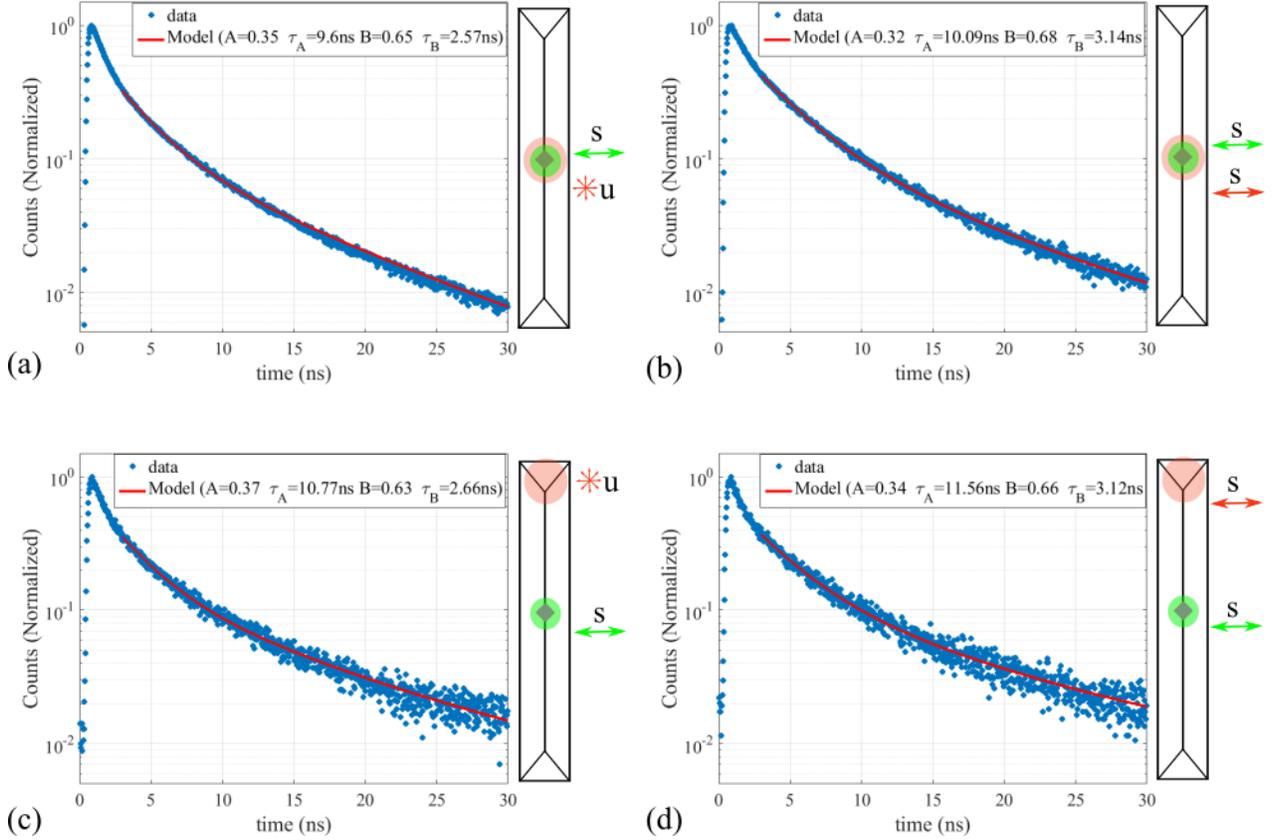

*Figure 9: Lifetime measurement data and a two-exponential fit for the fluorescence collected from the ND inside the VG. In sketches next to the graphs, green and red disks represent the excitation and detection spots, respectively, whereas the arrows represent the excitation (green) and analyzed signal polarization (red).*

Figures 8(a) and 8(b) present the confocal fluorescence image of the area of the VG containing the ND. The excitation laser power is, again, 100 μW. One can observe a significant decrease in the fluorescence counts obtained from the ND inside the VG when compared to the ND on gold-air interface. This could be due to the reduced intensity of the excitation laser inside the ND. In Figures 8(c) and 8(d), the fluorescence images are overlapped with a SEM image of the VG. The fluorescence images are recorded while focusing the excitation laser on the ND, and the fluorescence images in the middle and those at the VG ends are recorded independently to optimize the signal. The acquisition time for the fluorescence images in the middle is 1 s, whereas for the images at the VG termination mirrors is 60 s. The images suggest that the fluorescence from the ND couples to CPPs supported by the VG, which propagate along the VG and are out-coupled to free space from the termination mirrors of the VG. This sequence is observed for both excitation polarizations, as shown in Figures 8(c) and 8(d). The difference in the far-field images of the two termination mirrors is attributed to the presence of NDs near the VG termination mirror B [Figure 5(c)]. The emission pattern for both excitation polarizations is the same due to excitation of the CPP modes in both cases.

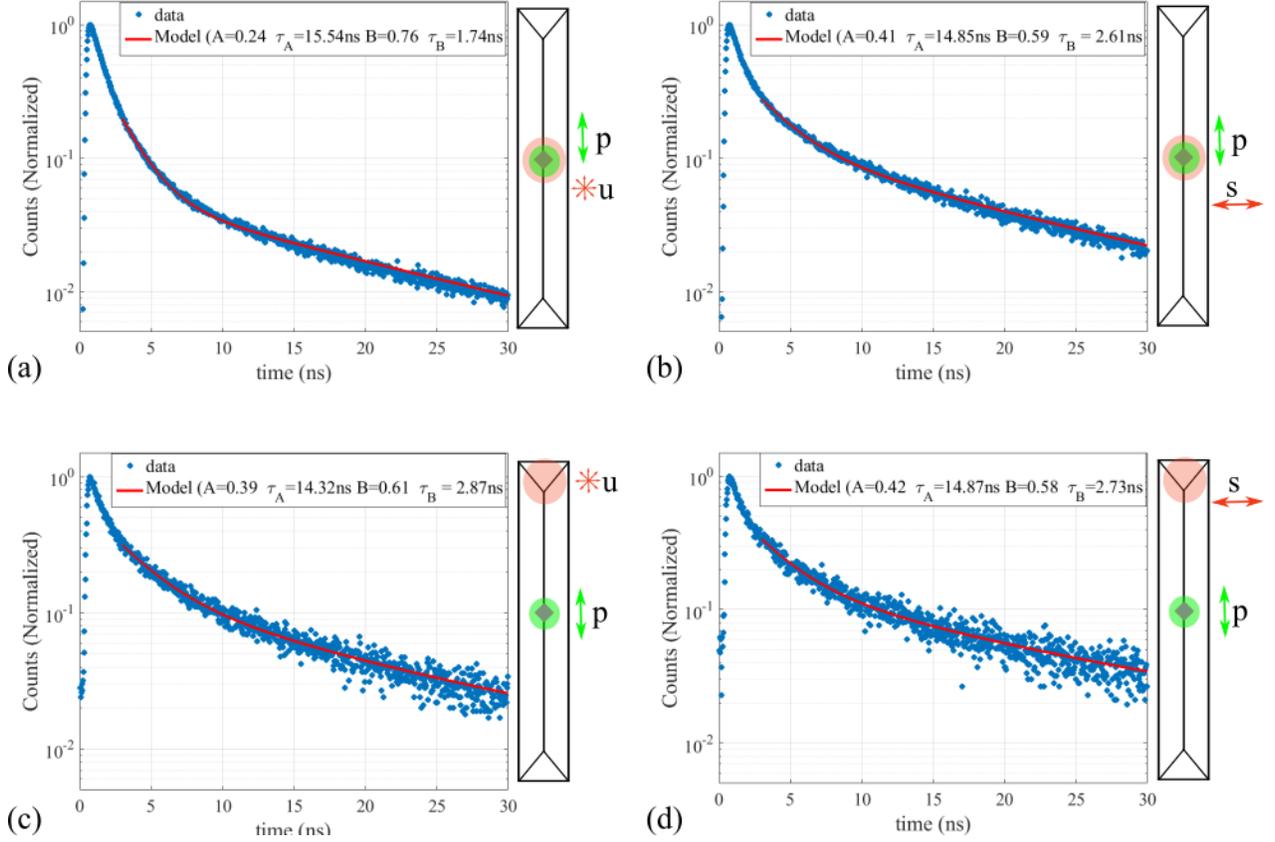

*Figure 10: Lifetime measurement data and a two-exponential fit for the fluorescence collected from the ND inside the VG. In sketches next to the graphs, green and red disks represent the excitation and detection spots, respectively, whereas the arrows represent the excitation (green) and analyzed signal polarization (red).*

The ND-VG system is further characterized by measuring the lifetime for different polarizations at the site of the ND, as well as at the VG termination mirrors. Figures 9 and 10 present the lifetimes of the coupled system when the excitation polarization is across the VG (*s* polarization) and along its axis (*p* polarization), respectively. In Figure 9, it can be observed that the values of lifetimes obtained at the site of the emitter as well as those at VG termination mirror A are very close. This suggests that the NV centers being probed with *s* polarization are the same, regardless of whether the fluorescent signal is probed at the end of the waveguide or at the site of the ND itself. Under these circumstances, NV centers with a non-zero projection of dipole moment along the excitation polarization are excited and coupled to the CPP mode of the VG. Previous studies of plasmonic modes in VGs indicate a more uniform distribution of the beta-factor across the profile when compared to cylindrical or wedge waveguides [33], which supports the assumption of a relatively uniform coupling of all the NV centers that are excited with *s* polarization. Furthermore, the fluorescence lifetime observed for the ND in this case is smaller than the fluorescence lifetime observed for the ND when it lied on the gold surface. The decrease for $\tau_A$ is by a factor of 1.32, and $\tau_B$ by a factor of 1.81. This is comparable to the lifetime change by a factor of ~2.44 observed when a single NV center in an ND lying on a gold surface is moved to inside a VG, in reference [28]. In figure 10, when the excitation polarization is along the VG axis, one can again observe that the values of lifetimes obtained at the end of the waveguide are very close to those at the site of the ND. However, if we compare the lifetimes obtained for the two orthogonal excitation polarizations, the lifetime in case of *p*-polarized excitation is longer than for *s*-polarized excitation. This can be attributed to the NV centers with dipoles across the VG axis, which couple more efficiently to the CPP mode of the VG. If we compare the fluorescence lifetimes of the ND when it is lying on the flat gold surface and when it is lying inside the VG and excited with *p*-polarized laser, the increase for $\tau_A$ is by a factor of 1.07, whereas for $\tau_B$ there is a decrease by a factor of 2.08. The slight increase of $\tau_A$ can be interpreted as a suppression of emission from dipoles aligned along the VG axis when compared to the ND lying on a flat gold surface. In Figure 11, we summarize all the lifetimesand amplitudes observed for the ND before and after its coupling to the V-groove. The errorbars represents uncertainty in the fitted parameters, that is, the lifetimes and amplitudes.

In Figure 12, we present two fluorescence spectra of the same ND inside the VG, one of which is collected at the site of the ND, whereas the other is out-coupled from the termination mirror A of the VG. In the former case [Figure 12(a)], the ND fluorescence spectrum is in superposition with fluorescence from gold, the spectrum of which is presented in the inset. In the latter case [Figure 11(b)], the contribution from gold fluorescence is negligible, since it does not couple efficiently to the CPP mode [28]. The zero phonon line of $NV^-$ centers can be observed in the spectrum that is measured at the end of the VG, which supports the argument above. The wavelength-dependent propagation loss of the CPP mode in the VG provides an additional source of modification of the observed spectra.

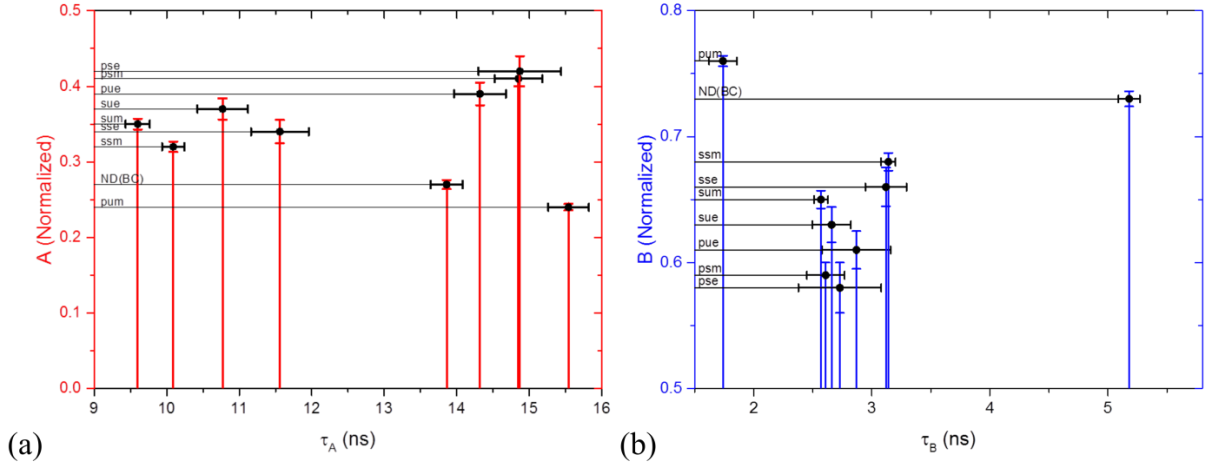

*Figure 11: (a) and (b) summarize the lifetimes and amplitudes observed in Figures 6(a), 9 and 10. ND(BC) denotes the nanodiamond before coupling to V-groove [Figure 6(a)]. In all the other symbols, the first letter denotes excitation polarization (s or p), the second letter denotes the analyzed polarization (s, p or u) and the third letter demotes the position of detection [middle (m) or end A (e)].*

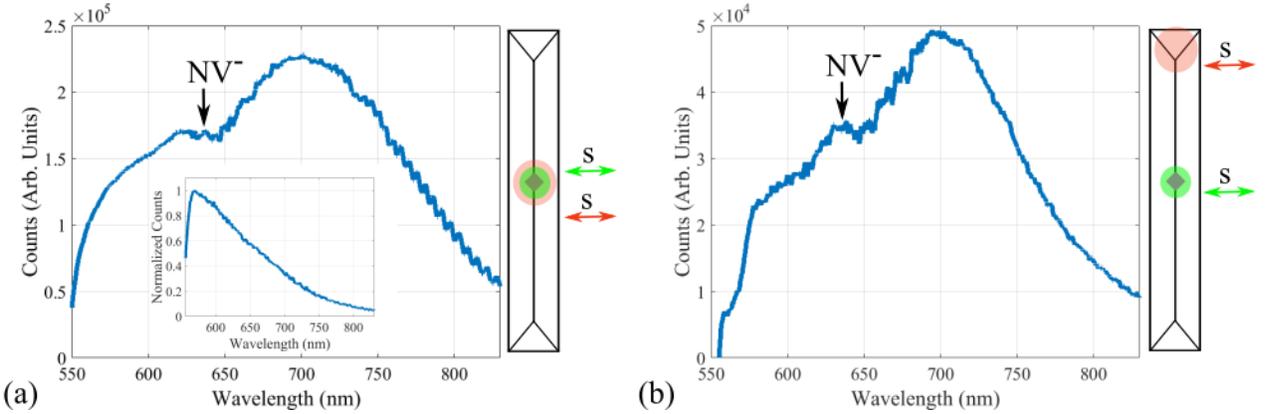

*Figure 12: (a) and (b) Spectra taken at the site of the ND and at one of the VG ends, respectively. The excitation and detection polarizations are across the VG (i.e. s-polarized). The inset in (a) shows the fluorescence spectrum measured from a gold film.*

We simulated the CPP mode that is supported by these VGs using finite-element method (COMSOL Multiphysics). The cross-section is obtained by a numerical process simulation in ATHENA, as has been described in [39]. The dielectric constant of gold is taken from reference [45]. In Figure 13(a), we present the distribution of the electric field of the CPP mode, calculated at a vacuum wavelength of 700 nm. The zoomed-in image in Figure 13(b) shows the distribution of electric field along with the possible position for the ND inside the VG. From the dependence of the CPP propagation length on the wavelength [Figure 1c(c)], the suppression of shorter wavelength is expected and is in support of the spectrum observed at the VG end.

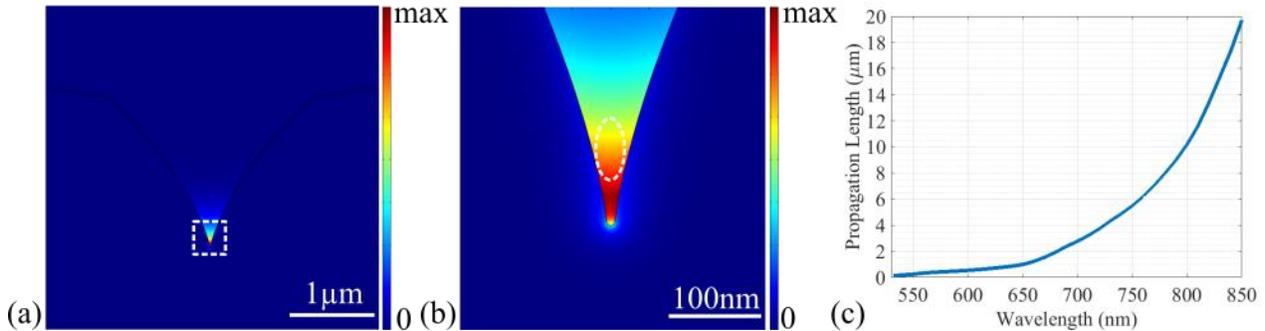

*Figure 13: (a) Simulated CPP mode in a V-groove. The dotted rectangle shows the area which is zoomed-in and shown in (b). The oval shape in (b) indicates a possible position for the ND. (c) The CPP propagation distance as a function of wavelength is presented.*

## 3. Summary

We have presented the coupling of NDs containing multiple NV centers to SPPs on a silver-dielectric interface and to CPPs supported by a gold VG. The coupling to SPPs is observed by emission from other NDs that are used for scattering the SPPs to the far field. The spectrum of scattered fluorescence is fitted to a model which suggests that the dependence on propagation losses as well as the height of the NDs, both of the source as well as the scatterer must be accounted for. We have also presented the results of coupling of an ND to the CPP mode of a VG. The coupling is demonstrated by the observation of fluorescence from the VG termination mirror while the ND pushed into the VG middle is excited. A lifetime change, when moving the ND from outside to inside the VG, is also observed for the ND fluorescence, which further confirmed the coupling. The change in spectrum with respect to the ND is explained as the change caused by the propagation loss plus the inefficient coupling of gold fluorescence to the CPP mode. The reported experiments demonstrate a way of incorporating NDs containing multiple NV-centers into plasmonic circuits as well as reveal various factors influencing the formation of fluorescence-excited SPP spectra.

## 4. Acknowledgements

The authors gratefully acknowledge the financial support from the European Research Council, grant No. 341054 (PLAQNAP). C.L.C.S. acknowledges financial support from the Danish Council for Independent Research (FTP Grant No. 12-126601).

19. A. Faraon, C. Santori, Z. Huang, V. M. Acosta, and R. G. Beausoleil, Physical Review Letters, 109, 033604 (2012), *Coupling of Nitrogen-Vacancy Centers to Photonic Crystal Cavities in Monocrystalline Diamond.*

20. J. Wolters, A. W. Schell, G. Kewes, N. Nüsse, M. Schoengen, H. Döscher, T. Hannappel, B. L. Öhel, M. Barth, O. Benson, Applied Physics Letters, 97, 141108 (2010), *Enhancement of the zero phonon line emission from a single nitrogen vacancy center in a nanodiamond via coupling to a photonic crystal cavity.*

21. J. Choy, B. Hausmann, T. Babinec, I. Bulu, M. Khan, P. Maletinsky, A. Yacoby, and M. Loncar, Nature Photonics 5, 738–743 (2011), *Enhanced single-photon emission from a diamond–silver aperture.*

22. S. Schietinger, M. Barth, T. Aichele, and O. Benson, Nano Letters 9, 1694–1698 (2009), *Plasmon-enhanced single photon emission from a nano-assembled metal-diamond hybrid structure at room-temperature.*

23. R. Kolesov, B. Grotz, G. Balasubramanian, R. J. Stöhr, A. A. L. Nicolet, P. R. Hemmer, F. Jelezko and J. Wrachtrup, Nature Physics 5, 470 (2009), *Wave–particle duality of single surface plasmon polaritons.*

24. A. Huck, S. Kumar, A. Shakoor and U. L. Andersen, Physical Review Letters, 106, 096801 (2011), *Controlled Coupling of a Single Nitrogen-Vacancy Center to a Silver Nanowire.*

25. S. Kumar, A. Huck, and U. L. Andersen, Nano Letters 13, 221 (2013), *Efficient Coupling of a Single Diamond Color Center to Propagating Plasmonic Gap Modes.*

26. S. Kumar, A. Huck, Y. Chen, and U. L. Andersen, Applied Physics Letters 102, 103106 (2013), *Coupling of a single quantum emitter to end-to-end aligned silver nanowires.*

27. S. Kumar, A. Huck, Y.-W. Lu, and U. L. Andersen, Optics Letters 38 (19), pp 3838-3841 (2013), *Coupling of single quantum emitters to plasmons propagating on mechanically etched wires.*

28. E. Bermúdez-Ureña, C. Gonzalez-Ballestero, M. Geiselmann, R. Marty, I. P. Radko, T. Holmgaard, Y. Alaverdyan, E. Moreno, F. J. García-Vidal, S. I. Bozhevolnyi, R. Quidant, Nature Communications 6, 7883 (2015), *Coupling of individual quantum emitters to channel plasmons.*
20

29. D. K. Gramotnev and S. I. Bozhevolnyi, Nature Photonics 4, 83 - 91 (2010), *Plasmonics beyond the diffraction limit.*

30. D. E. Chang, A. S. Sørensen, P. R. Hemmer, and M. D. Lukin, Physical Review Letters, 97(5), 053002, (2006), *Quantum optics with surface plasmons.*

31. D. E. Chang, A. S. Sørensen, P. R. Hemmer, and M. D. Lukin, Physical Review B, 76, 035420, (2007), *Strong coupling of single emitters to surface plasmons.*

32. A. V. Akimov, A. Mukherjee, C. L. Yu, D. E. Chang, A. S. Zibrov, P. R. Hemmer, H. Park, and M. D. Lukin. Nature, 450 (06230) 402, (2007), *Generation of single optical plasmons in metallic nanowires coupled to quantum dots.*

33. D. Martín-Cano, L. Martín-Moreno, F. J. García-Vidal and E. Moreno, Nano Letters 10 (8), pp 3129–3134, (2010), *Resonance Energy Transfer and Superradiance Mediated by Plasmonic Nanowaveguides.*

34. A. Gonzalez-Tudela, D. Martin-Cano, E. Moreno, L. Martin-Moreno, C. Tejedor, and F. J. Garcia-Vidal, Physical Review Letters, 106, 020501, (2011), *Entanglement of Two Qubits Mediated by One-Dimensional Plasmonic Waveguides.*

35. S. I. Bozhevolnyi, V. S. Volkov, E. Devaux, J.-Y. Laluet, T. W. Ebbesen, Nature 440, 508−511, (2006), *Channel plasmon subwavelength waveguide components including interferometers and ring resonators.*

36. D. K. Gramotnev and S. I. Bozhevolnyi, Nature Photonics 8, 13–22 (2014), *Nanofocusing of electromagnetic radiation.*

37. C. L. C. Smith, N. Stenger, A. Kristensen, N. A. Mortensen and S. I. Bozhevolnyi, Nanoscale, 7, 9355, (2015) *Gap and channeled plasmons in tapered grooves: a review.*

38. I. P. Radko, T. Holmgaard, Z. Han, K. Pedersen, S. I. Bozhevolnyi, Applied Physics Letters 99, 213109, (2011), *Efficient channel-plasmon excitation by nano-mirrors.*

39. C. L. C. Smith, A. H. Thilsted, C. E. Garcia-Ortiz, I. P. Radko, R. Marie, C. Jeppesen, C. Vannahme, S. I. Bozhevolnyi, and A. Kristensen, Nano Letters, 14, 1659−1664 (2014), *Efficient Excitation of Channel Plasmons in Tailored, UV-Lithography Defined V-Grooves.*
21